# Sowing 'Seeds of Doubt': Cottage Industries of Election and Medical Misinformation in Brazil and the United States

Amelia Hassoun[1], Gabrielle Borenstein, Beth Goldberg, Jacob McAuliffe, Katy Osborn

## Abstract

*We conducted ethnographic research with 31 misinformation creators and consumers in Brazil and the US before, during, and after a major election to understand the consumption and production of election and medical misinformation. This study contributes to research on misinformation ecosystems by focusing on poorly understood small players, or 'micro-influencers,' who create misinformation in peer-to-peer networks. We detail four key tactics that micro-influencers use. First, they typically disseminate 'gray area' content rather than expert-falsified claims, using subtle aesthetic and rhetorical tactics to evade moderation. Second, they post in small, closed groups where members feel safe and predisposed to trust content. Third, they explicitly target consumers' emotional and social needs. Finally, they post a high volume of short, repetitive content to plant 'seeds of doubt' and build trust in influencers as unofficial experts. We discuss the implications these micro-influencers have for misinformation interventions and platforms' efforts to moderate misinformation.*

## Introduction

Understanding why and how people consume, trust, and disseminate misinformation is crucial for developing effective strategies to combat its harms. This paper presents the results of ethnographic research with 31 misinformation creators and consumers in Brazil and the United States, before, during, and after the 2022 Brazil presidential election and US midterm elections. We examined the following questions:

*RQ1: How and why do people encounter, trust, and amplify misinformation?*

*RQ2: What tactics and signifiers of trust do misinformation creators use to influence others and amplify misinformation?*

*RQ3: How are consumers' trust heuristics and creators' tactics informed by the affordances and dynamics of online platforms?*

Existing literature often defines disinformation as explicit and deliberate fabrication (Bennett and Livingston, 2018; Damstra et al., 2021; Dan et al., 2021; Freelon and Wells, 2020), spread primarily by mega-influencers (Nogara et al., 2022). In this paper, we first argue for expanding disinformation studies' focus beyond political elites and popular superspreaders to 'micro-influencers' (<100K followers) who produce misinformation within what we term a 'cottage industry' of relatable, trusted peer-to-peer networks. By investigating these lesser-studied actors and locations of misinformation activity, our research shows how diffuse grassroots creation, sharing and engagement contribute to misinformation's spread and influence.

---

[1] Darwin College, University of Cambridge, Cambridge, United Kingdom. Email: ah2229@cam.ac.uk

Second, many researchers employ a strict definitional divide between 'true' and 'false' content, as well as between intentionally shared 'disinformation' and unintentionally shared 'misinformation' (Hameleers, 2022). Yet we found participants predominantly consuming, amplifying and creating what we term 'gray area' content (Krause et al., 2022) with multiple or ambiguous intentions and beliefs. A recent survey of 150 misinformation experts concludes that more research on these 'subtler forms of misinformation' is needed to develop 'better theories and interventions' (Altay et al., 2023:2).

In this paper, we define 'misinformation' as 'ideas that have yet to be the subject of a strong consensus of experts' (Uscinski 2023: 11).[2] We use the term 'gray area' to refer to the subset of misinformation not explicitly or entirely presenting information outside of expert consensus. This includes what we term 'bricolage content' (Levi-Strauss, 1966), where creators share content with expert consensus (like government data) but make additions (like putting it in a new context) that change its message. Content associated with 5G, stolen election, and the anti-vax movement were common in our participants' information ecosystems, but such conspiracy theories were only a subset of the misinformation we encountered.

With this definition, we avoid judging the veracity of content or the (often ambiguous) intentionality of creators. As Wardle (2023:38) argues:

> disinformation is distributed with the intent to cause harm, whereas misinformation is the mistaken sharing of the same content. Analyses of both generally focus on whether a post is accurate and whether it is intended to mislead. The result? We researchers become so obsessed with labeling the dots that we can't see the larger pattern they show.

In this paper, we instead focus on narrative and social patterns of misinformation sharing. Based on anthropological analysis of our participants' motivations, we find classifying content by veracity and intention less illuminating than analyzing the social and emotional benefits generated by sharing it.

---

[2] The data that support the findings of this study are openly available in the Open Science Framework (OSF) at http://doi.org/[TBD], reference number [TBD]. Access the recruitment screener for the specific election and medical ideas outside of expert consensus that we studied.

Our consumers-turned-creators evidence the complexity of intention and belief. All asserted that they had never intentionally shared false information. Instead, they sought to use information to connect with people and galvanize them around causes they deemed important, using strategies to avoid platform moderation when it obstructed their cause. We discovered that analyzing the veracity of individual content pieces was far less important to them than how important a cause (and its 'message') felt, requiring us to study their beliefs as dynamically constructed through practice rather than as statically held, always coherent realities. This shift helps us move away from binary definitions of actors as 'good' or 'bad' and more effectively trace misinformation sharers' often contradictory beliefs and multifactor motivations. We thus argue for the overlapping study of mis- and disinformation (e.g., Kapantai et al., 2020; Anderson, 2021), looking beyond typologies to analyze the social context and effects of this content on humans.

Third, we detail the underlying social and emotional motivations behind misinformation sharing, responding to researchers' calls for a 'more comprehensive picture of the emotional nature of misinformation' (Pasquetto et al., 2020:5; Kim and Chen, 2022). We find misinformation creation and consumption that fulfilled unmet emotional needs—for example, desire for recognition—aided its spread and influence, while participants often found content's veracity secondary or unimportant. We argue that researchers and institutions combating misinformation could be more effective by identifying and mitigating these unmet social and emotional needs, rather than primarily expending resources on classifying claims as false and debunking them. Misinformation sharing is a problem both due to circulation of potentially harmful information, and because it suggests unmet needs among a subset of the population. Our research is important because it argues that meeting the latter can help mitigate the harms of the former, without the partisan pitfalls of true/false classification (Uscinski 2023).

Finally, we found that repetitive exposure to similar misinformative messages (often in short-form content like memes or tweets) increased participants' misinformation belief and engagement. This finding contrasts with the prevalent epiphanic 'red pill' metaphor and related cultural imaginaries that individuals come to trust unorthodox or extreme views through singular, watershed moments (Stern, 2019; Madison, 2021). We also show how misinformation creators strategically focus on content quantity in attempts to foster an engaged community, planting many small 'seeds of doubt' across platforms.

We next situate our study in existing literature, detail our research methods and key findings, and discuss implications for future research and interventions.

**Sourcing misinformation: from big to small influencers**

Ethnographic explorations of misinformation creators since social media's emergence emphasize the role of democratized tools in crafting bricolage content—from enabling the production of GIFs to livestreams (Polleri, 2022; Woolley, 2023). As creation has gotten easier, misinformation has also become more subtle (Guess, 2020). Our study analyzes the emergent effects of this democratization on misinformation production and amplification.

Recent scholarship exploring how this democratization of production has led to misinformation proliferation typically diagnoses misinformation's severity based upon its reach, using metrics of impressions, views, or shares. Network modeling approaches that examine misinformation dispersion's nature and temporality often trace misinformation to public figures and influencers with large online followings (Nogara et al., 2022; Allcott et al., 2019; Allen et al., 2020). For example, the 'Disinformation Dozen' report found that 65% of COVID-19 misinformation on mainstream social media sites originated from 12 public accounts (Center for Countering Digital Hate, 2021).

Less attention has been paid to how smaller accounts contribute to misinformation creation and spread. Marketers and propagandists have popularized the use of micro-influencers (1,000-100,000 followers, sometimes called 'nano-influencers') by leveraging their localized influence and ability to build a loyal audience with higher levels of trust and engagement (Maheshwari, 2018; Conde and Casais, 2023).

Within misinformation studies, we build on ethnographic research into micro-influencers paid to promote propaganda due to their 'localized, relationally potent effect' (Ong and Cabanes, 2018; Woolley, 2022:119). Micro-influencers leverage self-disclosure, perceived authenticity, and familiarity to build 'parasocial relationships' that deeply engage their audience (Harff et al., 2022; Stehr et al., 2015). Given the advantages micro-influencers have in building trust, we argue that the effects of micro-influencers on misinformation belief and amplification is understated in a literature focused on popular influencers and institutions.

We propose that micro-influencers influence misinformation spread in part because information shared by 'regular people' increases others' susceptibility to it (Anspach, 2017). People typically perceive sentiments shared by relatable individuals, rather than celebrities or sponsored influencers, as more credible because they seem less biased by profit motives (Hassoun et al., 2023). Their content format feels familiar, mirroring the unpolished, personal content people encounter in social media ecosystems (Anspach, 2017).

Another structural reason for micro-influencers' relatively high engagement is that they typically operate in trusted, closed networks that resemble word-of-mouth communication. Their '"atoms" of propaganda…rocket through the information ecosystem at high speed powered by trusted peer-to-peer networks' (Wardle, 2017). Burgeoning private messaging services have increased the speed and relatability of these 'atoms' (Rossini et al., 2020). Despite the proliferation of research on misinformation influencers across platforms, this scholarship has largely discounted the network effects of micro-influencers specifically. Our research suggests that this increase of micro-influencers using closed, peer-to-peer networks may meaningfully impact misinformation belief and sharing.

**Misinformation: from explicit, intentional fabrication to 'gray area' content**

A series of 2020 Reuters Institute studies found that most online COVID-19 medical misinformation sampled was not purely fabricated (Brennen et al., 2021). Rather, most involved 'various forms of reconfiguration where existing and often true information [was] spun, twisted, recontextualized, or reworked.' The most common reconfiguration was 'misleading content' combining expert consensus and non-consensus information (what we call 'bricolage' content). For example, one 'very widely shared post' gave 'medical advice from someone's uncle, combining both medical expert and non-expert information about how to treat and prevent the spread of the virus' (Brennen et al., 2021). Such bricolage content drives more social media engagement and sharing, in part because it is often 'emotive' and 'taps into base emotions like fear or outrage' (Marchal et al., 2019).

Most misinformation our participants produced and shared falls under this 'gray area' category. Because gray area content mixes expert-verified and unverified content or avoids making verifiable claims entirely, it is difficult for moderators to automatically detect, develop policy around, and moderate. Participants often spread this hard-to-detect-and-moderate content in direct response to the threat of deplatforming or demonetization, an example of creators 'evolving along with the information landscape' (boyd, 2017). We show how this content requires different misinformation classification and moderation approaches, because its creators employ tactics like rhetorical questions and implied correlations instead of standalone expert-falsifiable claims.

Misinformation researchers often approach belief ontologically, systematically classifying content into belief systems like QAnon or miracle cures (Kapantai et al., 2020). Researchers have also sought to label actors as "good" or "bad" based upon whether actors think that the information they share is false and whether they share it for personal gain (Hameleers, 2022). Drawing on anthropological literature demonstrating that belief is made and sustained through community-based practice, we argue for shifting research questions towards how individuals acquire and sustain belief (Hassoun et al., 2023; Luhrmann, 2020; Deeb 2011). We found misinformation belief was built and sustained through community-based practices of sharing; belief did not necessarily precede sharing (Ren et al., 2023; Fountain, forthcoming).

Moving beyond these classifications allows us to focus on the well-documented array of harms resulting from misinformation. At the individual level, misinformation can motivate people to consume harmful substances believed to be miracle cures, decrease vaccination intent, and avoid medical experts, all which increase health risks (Loomba et. al., 2021). Research into the correlation between misinformation belief and emotional wellbeing finds significant associations between misinformation belief, depressive symptoms, and low life satisfaction ratings (Perlis et al., 2022). Additionally, the repetition of misinformation narratives containing stereotypes amplifies racist, misogynist, xenophobic, and transphobic tropes (Phillips and Milner, 2021; Polletta and Callahan, 2017). Parallel literature details misinformation's corrosive effects on democratic institutions and societies, due to its contributions to affective polarization and anti-deliberative effects on discourse (Tucker et al., 2018; McKay and Tenove, 2020). Because gray area misinformation creates such challenges for moderation and detection, these harms become amplified and harder to combat.

## Methodology

We conducted the first phase of a two-year study in Fall 2022, employing ethnographic methods to longitudinally analyze why and how people consume, amplify, and create medical and election misinformation. Being in the field before, during, and after Brazil presidential elections and US midterm elections allowed us to study the dynamic relationship between political misinformation, belief, and action—culminating in the storming of the Brazilian Congress on January 8th, 2023. Existing misinformation research is highly US-centric, with South America being the least studied (3.8% of studies) region in the world and 'only 7.6% of studies analyzing US data along with those from other countries' (Seo and Faris, 2021:1166).

We chose to study both creators and consumers because existing research focuses primarily on misinformation *consumption*. Studying creators allowed us to understand what motivates individuals to move from passive consumption to active amplification and creation.

**Participants & Sites**

We conducted semi-structured interviews and ethnographic participant-observation with 31 participants aged 18-67 who regularly created, amplified, and/or consumed misinformation (Table 1). We also attended 3 misinformation-spreading events that participants engaged with. In Brazil we visited local events, like a Sunday church gathering that served as a key misinformation source for its community. In the US, multiple participants invited us to ReAwaken America, a 5,000+ person conference series that National Public Radio describes as 'part QAnon expo and part political rally' (Hagen, 2022).

Researchers natively spoke English (US) and/or Portuguese (Brazil). Participants spanned education levels and political affiliations.

**Table 1**

*Participant Information*

|  | Total N | Men (self-ID) | Women (self-ID) | Urban | Rural | Misinformation Type | | |
|---|---|---|---|---|---|---|---|---|
|  |  |  |  |  |  | Medical (only) | Political (only) | Both |
| Site |  |  |  |  |  |  |  |  |
| Brazil | 16 | 9 | 7 | 12 | 4 | 2 | 4 | 10 |
| US | 15 | 6 | 9 | 10 | 5 | 5 | 4 | 6 |
| TOTAL | 31 | 15 | 16 | 22 | 9 | 7 | 8 | 16 |

**Recruiting & Incentives**

We identified relevant public social media groups, subreddits, and chat app channels. Disclosing ourselves as researchers, we built rapport and used direct messaging to provide study information and a recruitment screener. Participants received $100/hour (US) or 250 reais/hour (Brazil). Our study underwent human subjects ethics review and participants gave informed consent. All personally identifying information is omitted.

**Research Methods**

We conducted three-part, 6-8 hour online and offline ethnographic interviews with each participant. We chose this methodology to deeply understand how participants encounter, share, and create misinformation within their everyday lives and to develop a broader understanding of how on- and offline misinformation ecosystems connect.

*Remote Semi-Structured Interview + Observation*

We began with a 3-hour Zoom session examining participants' online lives and behaviors, combining interview, remote screen-sharing, and observation as they led us through their online ecosystems. Prompts used to investigate participants' beliefs and behaviors were informed by prior ethnographic studies and open-ended. Researchers mirrored respondents' mental models and language, letting participants guide conversation.

*In-Person Semi-Structured Interview + Observation*

We then conducted 3-hour sessions in participants' homes and social spaces to understand the relationships between their online and offline information ecosystems.

*Semi-Structured Interviews with Secondary Participants*

During in-person sessions, we asked participants to introduce us to 1-2 important people in their lives. We had 1-hour conversations with a partner, close friend, or community member who helped contextualize participants' beliefs and online behaviors in their social ecologies.

**Analysis Methods**

Grounded theory guided data analysis (Charmaz, 2006), with researchers documenting images, video, and field notes during participant-observation. Researchers collaboratively performed open coding, clustering, and thematic analyses (Saldaña, 2021).

**Limitations**

We chose to study participants ethnographically to give in-situ observational detail and qualitative explanation to a phenomenon most commonly studied using lab-based experimental or computational methods (Seo and Faris, 2021). Follow-up surveys on a representative sample would help explore whether our findings have broader applicability. Self-reported data limitations include self-censoring, recall challenges, and social desirability biases. We sought to analyze gaps between what participants say and do by cross-referencing semi-structured interview data with digital artifacts (e.g., search and message histories) and screen-sharing observation while participants navigated their digital ecosystems.

## Findings

In this section, we detail four key findings from our ethnographic research (Table 2).

**Table 2**

*Key Findings*

| Tactic | Action | Rationale |
| --- | --- | --- |
| 'Gray Area' Content | Imply rather than state (mis)information | Avoid detection and moderation |
| Micro-Influencing | Post in small, closed groups | Create a sense of intimacy to gain deeply committed, trusting followers |
| Emotional Targeting | Speak to and satisfy unmet social and emotional needs | Create dependence on misinformation community |
| Quantity over Quality | Plant small 'seeds of doubt' repeatedly across platforms | Subtly build belief over time |

**'Gray Area' Content**

Creators used gray area content to avoid moderation, which many experienced after sharing what platforms deemed misinformation. We detail four common forms: bricolage content, personal testimonials, pseudo-scientific jargon, and strategic questions.[3]

*Bricolage Content*

Creators posted bricolage content to suggest causation without explicitly claiming it. Tina (52, US) connected images of 5G towers and children with cancer, urging consumers to draw their own conclusions: 'I wanted the videos to stay up on YouTube, so I just let the images speak for themselves.'

Brandon (41, US) created and shared a graph, using real government data, to imply that 'excess deaths' during COVID were caused by COVID-19 vaccinations rather than the disease itself

---

[3] Other creator tactics to avoid moderation included using satire, opinion, and neologizing, as Murilo (43, BR) explained: 'You can't say "Alexandre de Moraes", so I'll say "Xandex Xandovski" when he does something bad. Then I don't fall into the algorithm.' Creators were hyper-aware of 'the algorithm' as an (unfairly) punitive force.

(Figure 1). 'The data speaks for itself,' his caption read, 'just look at the graph.'

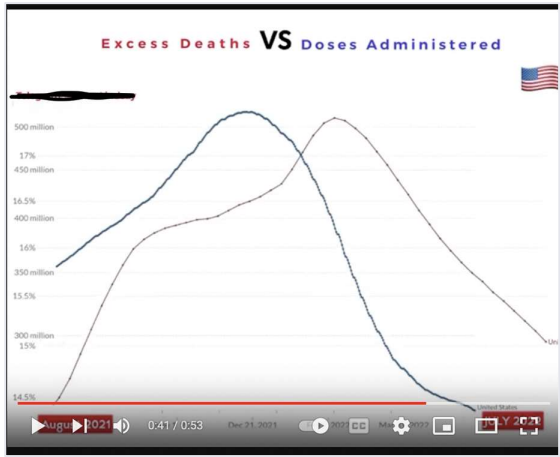

Figure 1: Brandon (41, US) bricolages government data on excess deaths during COVID with vaccine doses to imply causation.

*Personal Testimonials*

Creators also shared personal testimonials that were emotionally compelling and difficult to refute. Tina (52, US) narrativized her health struggles to sell anti-5G products. Nadia (34, US) shared her mother's illness story to warn against trust in clinical experts:

> She was put on narcotics…I was watching her bedridden and no longer coherent. I went through a phase where I was grieving her while she was physically still alive, but then I started learning about plants and natural medicine…eventually it helped her pain…She got off all her narcotics completely. I said: Plants over pills.

By sharing testimonials (Figure 2), a group followed by participants promoted ivermectin treatment and anti-vaccine beliefs, planting seeds of doubt about institutional medicine while avoiding moderation. These testimonials encouraged consumers to reject drugs/vaccines and avoid medical experts, causing potential harm.

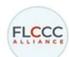

Figure 2: Personal testimonial from a wife whose husband died while hospitalized for COVID.

*Pseudo-scientific Jargon*

Creators often used technical, pseudo-scientific jargon to portray expertise and credibility, stylistically mirroring scientific experts' use of terminology and evidence. Nadia (34, US)'s critiques of big pharma often called out particular drugs, chemical compounds, and biological processes that exceeded her followers' common knowledge: 'there are a lot of people out there just regurgitating information, I want my followers to know I read all the scientific literature and know what I'm talking about at the cellular level.' Consumers like Alan (27, US) remarked how such 'technical lingo' makes content seem credible (Figure 3). When a creator showcases specialized knowledge, it builds his trust in their expertise–that '[the creator] has done all the research for me.' These consumers replace trust in scientific institutions with trust in creators as their go-to sources (Hassoun et al., 2023) for information.

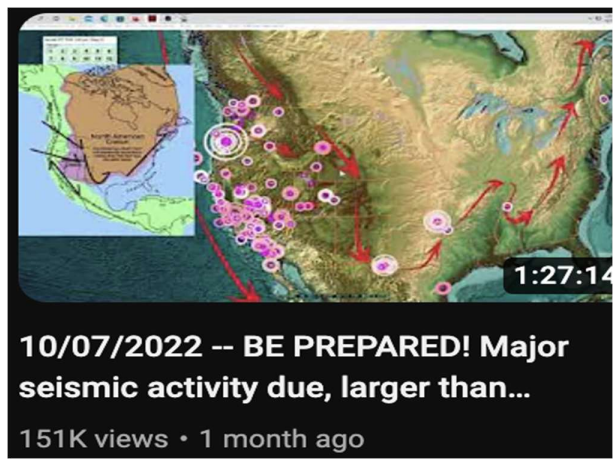

Figure 3: Alan (27, US) feels the pseudo-scientific jargon in this video means its creator is a credible 'expert.'

*Strategic Questions*

Creators used pointed, provocative, rhetorical, and other strategic questions to challenge expert consensus without explicitly making expert-falsified claims, thereby evading moderation. Nadia (34, US) frequently pointed consumers to parallels between the COVID-19 vaccine and groupthink. For example, she tweeted: "The jab is the real-life scenario your mother warned you about. 'If your friends all jumped off a bridge, would you?' Now we know who would jump."

Tina (52, US) encouraged consumers to doubt vaccines (Figure 4) by asking questions implying causation: 'healthy individuals losing their lives. Is it the vaccine? Well, interesting that they've all taken it.'

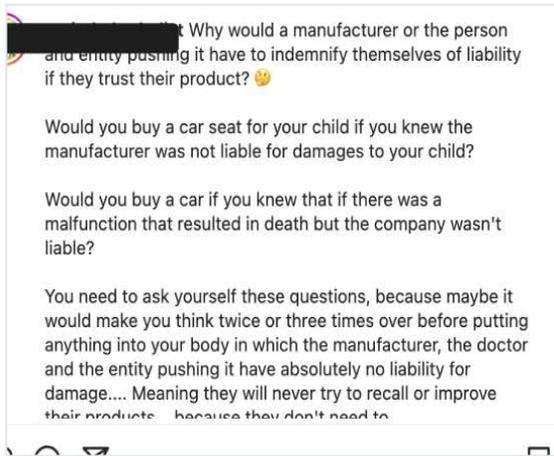

Figure 4: Tina (52, US) uses provocative questions comparing vaccines to car seats to raise stakes, highlight risk, and discourage vaccination.

Creators also built trust by giving consumers strategic questions to type into search engines. These creator-suggested 'search queries with an ideological dialect or bias' (Tripodi, 2022) led consumers into 'data voids' filled with creators' (and their communities') search engine optimized content (Golebiewski and boyd, 2018). For example, when Ted (60, US) was told to search 'What is the World Economic Forum's 2030 Agenda?,' results led him to other conspiratorial queries (Figure 5). This built his community-driven suspicions about WEF. 'If [creators] tell you to look it up for yourself, and they tell you where you can go to do it, I start to believe that.' he said. 'And I do go there and look it up.'

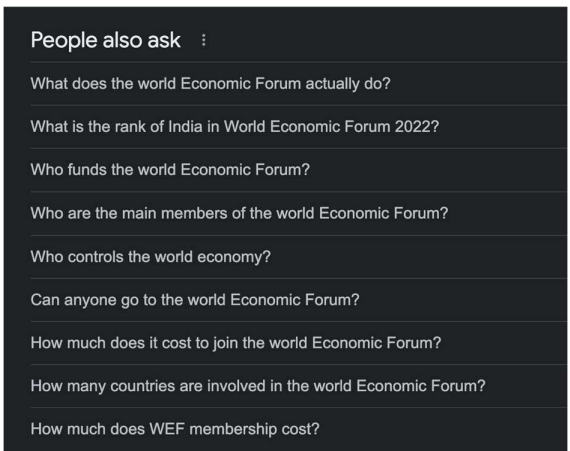

Figure 5: Ted (60, US)'s 'People also Ask' suggestions.

**Micro-Influencing**

Our participants' information ecosystems were primarily shaped by micro-influencers speaking to small, often closed, online groups, rather than mega-influencers. Because consumers encountered misinformation content in spaces they felt safe, it often subtly blended into their feeds through the evasive tactics detailed in the previous section.

*Small, Local, Closed Groups*

Participants typically encountered misinformation within trusted spaces, often from sources already familiar or familiar-seeming. Many first encountered and consumed misinformation from friends and family via direct messages. Conceição's (67, BR) son sent her videos: 'He recommended this YouTube channel. He said Mom, look at Luciano, he is also from Caxias.' I started following him and I liked it. I watch it every week.' These personalized recommendations drew participants to new (mis)information sources and communities.

Participants demonstrated a clear preference for information from local communities. Dave (58, US) joined his Concerned Doctors chapter (Figure 6) because they offered hyper-local news and opportunities: 'They send out a weekly newsletter with the latest on what they are finding, and they share in-person events where I've made a lot of friends.'

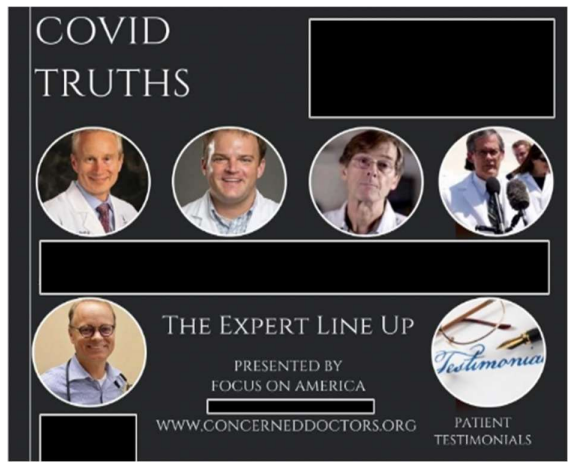

Figure 6: Local Concerned Doctors chapter spreading misinformation via testimonial and experts.

Paulo (64, BR) relied on a church WhatsApp group for information:

> We created this group ['Faith, Politics, and Economy'] so we don't have to share information individually. These people [have been] friends for a long-time. The majority are friends from the church, so I know them in real life.

Early misinformation encounters also came from unofficial 'experts' (e.g., influencers, podcasters) who participants trusted. Lorenzo (34, BR) first learned about election misinformation through his cryptocurrency Facebook groups and favorite YouTube channels, which felt like safe spaces to learn new ideas and find community. He joined in 2015 when scared by inflation in Brazil and slowly acquired election conspiracy beliefs: 'I followed the YouTube channels to learn more about finance and bitcoin, but ended up joining news channels that led me to conspiracy channels.'

Creators sometimes repurposed closed groups to introduce members to misinformation in an already safe space. Susanna (43, BR)'s Telegram group 'Doctors for Life' was renamed to 'Politics and Health' and later spread election misinformation as 'Geopolitics SOS Army.'

Some consumers felt empowered to become creators in these trusted circles. Susanna (43, BR) started consuming (mis)information months prior to the 2018 election when friends added her to WhatsApp groups. She proceeded to lead 60+ WhatsApp groups during the election and proudly said she is a trusted source: 'People know me…On WhatsApp, when anything happens in politics, I get many messages saying 'Is it true?' 'Did you see this?' Now, I am a reference.'

*Personal Relatability*

Because of this personal, recommendation-based information circulation, creators felt that grassroots personal relatability contributed significantly to their success. Nadia (34, US) shared her '6 figure formula' for building a loyal online following:

> You need your healing story…you will show your story time and time again, it makes you a real person and relatable. That's where you build your credibility. My success all comes down to my story. Hearing about my experiences—how I hit rock bottom and healed myself, helped heal others naturally, made people want to learn more from me. It's how I grew my presence online. I never used to use social media before but it's easy once you figure out how to harness your story.

Tina (52, US) similarly attributed her success to relatability: 'My brand's success comes from my story. It makes people trust me—I've been through it and I'm better now.'

*Belief and Incentives*

Creators both asserted belief in the misinformation they spread and had clear incentives to amplify it. Tina (52, US) believed she had the solution to 5G and made six figures selling it. Murilo (43, BR) believed Brazil's election was rigged and used it to gain fame. Brandon (41, US) believed the COVID vaccine kills and spread the word to build his stature as a good Christian.

Amplifier participants passionately and firmly asserted that they had never shared false information. Ted (60, US) said: 'Everything is something that I have checked out for myself, or I know to be true. I would never post misinformation intentionally. I don't think it's ethical.' Allison (47, BR) asserted: '[5G] is not conspiracy theory, it's reality theory.' Fernanda (48, BR) read about Russian disinformation tactics (Figure 7) and accused Bolsonaro's political opponents of using them—but vehemently asserted that she never had.

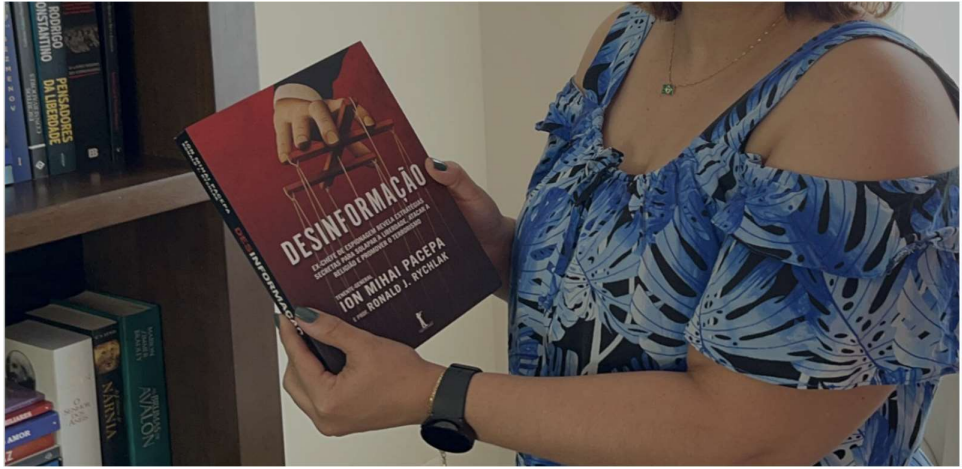

Figure 7: Fernanda (48, BR) showed us a book to demonstrate that she knows disinformation when she sees it.

**Emotional Targeting**

When participants encountered misinformation, they tended to engage and share when it satisfied key emotional needs: countering loneliness, feeling valued as an expert, alleviating fears, and venting frustrations. Creators recognized and fed these needs.

*Loneliness*

Misinformation sharing offered participants real and imagined communities that abated feelings of loneliness. For Brandon (41, US), online and in-person anti-vax events provided a sense of community belonging during COVID: 'When Plandemic lockdowns started, we were so isolated in Texas. I said to [my wife] Delilah, we gotta find some like-minded people to engage with. When I learned about ReAwaken America, I knew we needed to be a part of it.'

Livestreaming enabled viewers to quickly build online communities, finding real-time emotional connection and validation. Participants attended livestream watch parties or coordinated viewings to collectively react to events (Figure 8). Creators often held these at strange hours to build a special sense of community while avoiding moderation. Fernanda (48, BR)'s favorite livestream promoted election fraud. She enjoyed the community it fostered: 'Our livestreaming is more like a chat. It's not very professional or formal. The people talk as if it were a WhatsApp conversation. They participate a lot.' Restricted access to pages and events also gave participants both a sense of belonging and special importance.

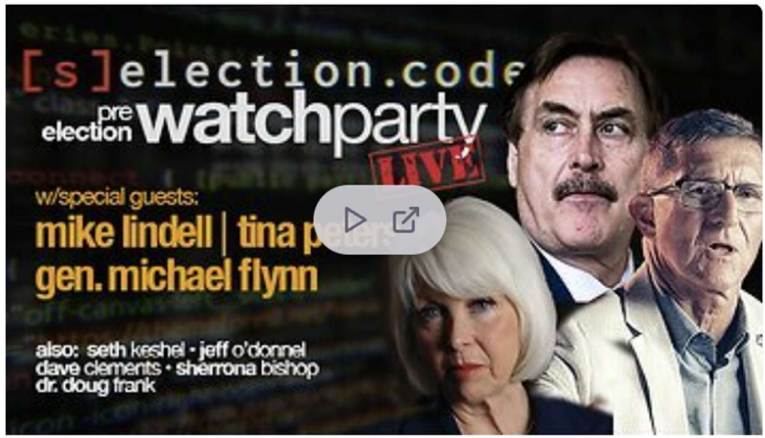

Figure 8: Dave (58, US) co-watches election coverage live on Rumble using a link he found on Mike Lindell's Truth Social page.

*Feeling Valued as an Expert*

Misinformation sharing provided participants an antidote to obscurity. Many participants felt undervalued for their abilities or intelligence. Online misinformation communities provided positions of power to those who felt powerless. Susanna (43, BR) appreciated that an election misinformation community recognized her intellect and experience: 'I'm not rich, I have some dignity from my work. But when I joined the community, I started to have access to people I wouldn't have imagined. Like big businessmen, talking as equals, talking about politics.'

Fernanda (48, BR) had law and business degrees but practiced neither profession. She found professional recognition and a sense of purpose through an election misinformation community: 'I was one of the people who used to comment...one day they called me to do [a livestream]. I never planned it. In life there are moments you have to make a choice: Either you keep quiet, carry on with your life and let the world fall apart, or do something.' Alan (27, US) felt validated when people responded to and re-shared his Reddit and Telegram posts: 'It tells me I'm going in the right direction. It's something I can use to see how much people engage with [my content], how important it is for other people, how current it is.'

*Fear*

Fear motivated misinformation sharing. Dona (62, BR) described an election fraud video shared with her that she forwarded to others: 'This was after the election, this woman found a box full of ballot paper on the street...I thought it was very serious. I don't know if it's true or not. Let's leave this to God, he knows.' This urgency made fact-checking secondary for Dona and other amplifiers, making a fearful 'just in case it's true' feeling override veracity concerns.

Creators marketed fear and sold solutions to give consumers a sense of control in uncertain times. Tina (52, US) originally felt sick and 'suspected it was from radiation. But no doctor had any answers, and that's the scariest part.' She found answers in 5G conspiracy: 'It all just keeps getting worse, too–look at how many towers there are! I knew I had to figure out something to do.' We saw at ReAwaken America how Tina eventually made a living selling her fear–and a solution: 'I touch and interact with at least 2000 people at each event, reading their radiation levels and showing them how bad 5G has gotten…I give them the products I wish I had.'

We met many similar vendors at ReAwaken. Brandon and Delilah (41, US) sold billboard advertisements to 'spread the word' and market fear about the COVID vaccine. Delilah said: 'The fact is they're coming after kids. People are dying from these shots and now they're mandating children get vaccinated." Brandon echoed: "If I don't [act], it's like blood on my hands. You have to do something even if you're scared. You need to speak up, speak out.' Others sold products like vitamins to protect against the perceived negative impacts of vaccines and 5G.

*Judgment-Free Venting*

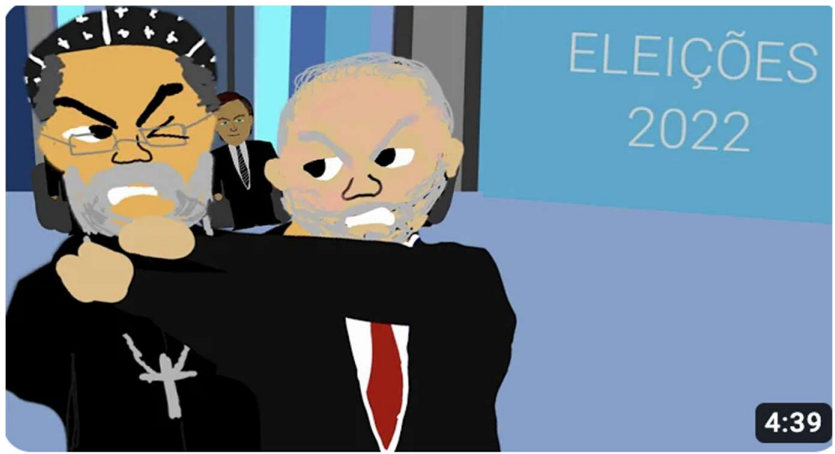

Figure 9: Creators like Murilo (43, BR) made 'satirical' political videos to provide consumers with emotional outlets and misinformation beliefs while also evading moderation, as humor makes creator intention hard to determine.

Finally, many participants liked misinformation communities because they provided anonymous or safe, judgment-free zones to vent frustrations (Figure 9). Reddit provided Kara (34, US) with an alternative to 'normie' platforms, where: 'I don't feel like I can fully share my experiences and thoughts…and it's not just about censorship and posts removed. I feel this strong sense of judgment…You should be able to say things without automatically being labeled anti-vaxxx with three x's.' Ted (60, US) frequently reshared posts on CloutHub and MeWe to vent: 'This affords you an avenue of relief. Otherwise I would just keep it all bottled up. I would explode.'

These emotionally-driven reasons for misinformation sharing indicate potential for interventions that meet these emotional needs through alternative means, but also demonstrate the deep hold misinformation communities can have. Online gardening communities initially pulled Allison (47, BR) away from misinformation communities, as they ameliorated his feelings of loneliness and obscurity. But the election generated fear and frustration, driving him back to misinformation communities that addressed those emotions.

Emotionally unsettling events catalyzed more active engagement with trusted misinformation spaces. After a significant event—from elections to COVID—Ted (60, US) turned to the 98 pages he followed on CloutHub or to TruthSocial: 'As soon as something happens, someone is talking about it there' (Figure 10). Following the election, Fernanda (48, BR) returned to Jovem Pan and Bolsonaro's pages, as well as familiar podcasts and livestreams where she previously found election misinformation. These spaces became her go-to sources: 'Everyone's here. I find it easier to find things around here, you see? Bolsonaro's page is the first place that I look.'

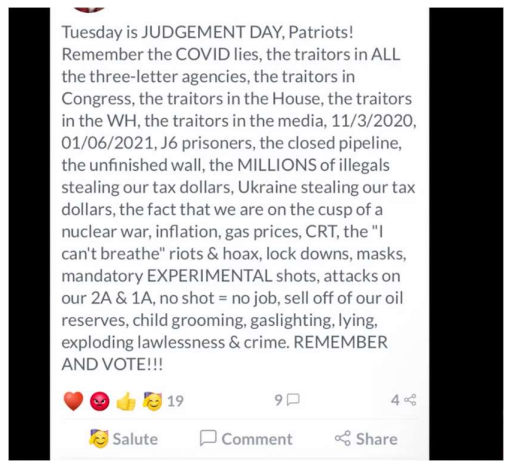

Figure 10: Ted (60, US) shares an emotionally charged post connecting myriad misinformation and conspiracy theories.

Emotionally unsettling health concerns also drove participants to become 'miracle cure' misinformation consumers and creators. Nadia (34, US) explained: 'I was desperate for another way. That's when I started researching.'

**Quantity over Quality**

Misinformation creators employed the tactic of repetition, planting many 'seeds of doubt' in consumers' trusted spaces over time. Nadia (34, US) recognized that to gain new followers, (mis)information quantity was more important than quality. She then increased her posting frequency and posted similar content across multiple platforms (Instagram, TikTok, Twitter, Facebook, Telegram, Bitchute, Rumble). After her posts were removed or downranked, she learned to evade each platform's misinformation policies:

> When the rhetoric around COVID intensified I started realizing some platforms are going to be easier to get the word out than others…I stopped talking about the vaccine and COVID. I just started planting seeds about the forms of corruption.

This approach enabled Nadia to continue amassing followers even when she encountered moderation: 'When I initially had my account disabled I was freaking out, this is my business account! But then I made a TikTok and I jumped onto Twitter. I will hop on everywhere to get the word out.' Creators like Nadia recognized and capitalized on word-of-mouth re-sharing: 'I have countless messages of people being like: "I shared your video with my family who is now doing this." Her tactics gained her many followers on Instagram (67K+), TikTok (44K+), Twitter (14K+), YouTube (2K+) and Telegram (5K+).

These cumulative seeds of doubt worked to build trust and belief with several of our participants. Allison (57, BR) could not remember how he came to believe in 5G: there was no singular, red-pill moment, just an accumulation of moments as he repeatedly encountered videos, posts, and articles. Participants like Allison granted limited thought to misinformation seeds of doubt in initial encounters: 'I heard about an experiment in which birds died after landing on a 5G wire. Is this true or not? I don't know.' Over time doubt increased: 'I heard it is bad for your health…It is bad for your brain. It is really scary!'

We found that short posts on Instagram, TikTok, and WhatsApp were particularly effective at planting many seeds of doubt for participants within a short timeframe. Long-form lists with links to additional information similarly inundated consumers with a sense of accumulating evidence. At ReAwaken America, presenters shared long documents containing questions and misinformation evidence using SMS, Google Drive, and Dropbox (Figure 11).

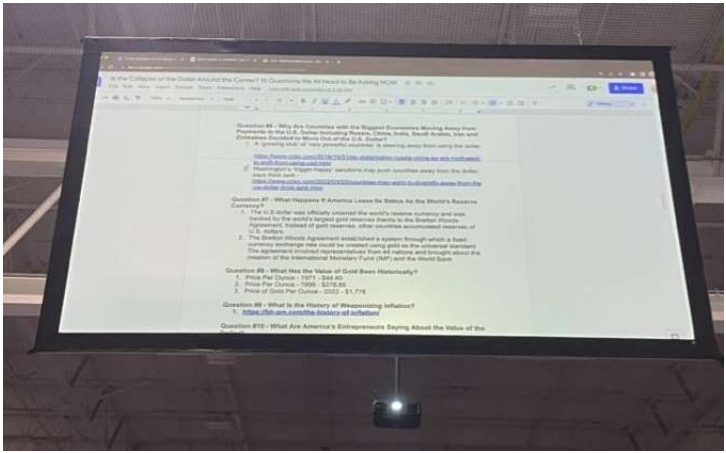

Figure 11: ReAwaken America presenters shared long pages of 'evidence' with audiences.

Repeated algorithmic recommendations also drove participants to misinformation during their routine online activity in trusted spaces, giving them a similar sense of accumulating evidence. Alan (27, US) used suggested videos to discover music, but the recommendations led him to misinformation spaces: 'My YouTube used to give me music channels. Now it is all politics and what is going on in the world. The algorithms start to dominate my sidebar.' Ted (60, US) received podcast suggestions through algorithmic recommendations in the streaming app he already frequented: 'The podcasts start popping up everywhere. And you're like, man, I never even knew this was going on! It's really waking people up.'

In the next section, we discuss these findings and their implications for future misinformation research and interventions.

## Discussion

Our findings demonstrate the need for research on the effects of repetitive, 'gray area' content prevalent for our participants and prolific in today's information ecosystems. Methods that sequentially analyze the harmfulness of standalone pieces of content miss the cumulative effects of misinformation sharing. Participants found the accumulation of many, often short online messages reinforcing the same point to be persuasive, regardless of the content's production quality (or veracity). In fact, creators asserted that their posts' quality was less consequential for their success than frequency, authenticity, and customization for each platform.

**Micro-Influencers as Unofficial Experts**

This proliferation of low-quality content coincides with shifts in people's trust heuristics. Polls find declining trust in institutionalized authority and elite expertise, with a corollary decline in trust heuristics like (inter)national institutional approval (Gallup and Knight Foundation, 2020). Many people increasingly use signals of familiarity as trust heuristics, like our participants who prioritized local influencers and groups. Given legitimacy need not be conferred by traditional markers of institutional authority, creators can confer credibility through personal testimony, entrepreneurship, and deep engagement with community members. This shift enables the decentralization of misinformation creation to smaller creators in what we term a 'cottage industry of misinformation.'

We observe a complementary trend emerging from declining institutional trust: the rise of the 'unofficial expert'. Unofficial experts project an image of possessing information not shared or explained by mainstream institutions (Jigsaw, 2022). Sharing such information positions these 'experts' as truth-tellers defying traditional authorities and institutional information sources. This is distinct from the 'fake expert' phenomenon, whereby individuals stylistically imitate authoritative information sources like news outlets to spread misinformation (Cook et al., 2017). 'Unofficial experts' need not repurpose the trust heuristics of mainstream media like formalwear or high production value videos; their credibility comes from being ordinary people with seemingly extraordinary knowledge (Hassoun et al., 2023). Our participants were drawn to and modeled such unofficial experts.

**Misinformation meets emotional needs**

By serving as unofficial experts, creators often met their own and others' emotional needs. Creators found validation from gaining followers and engagement or becoming a sought-after source of information for a community, mitigating the feelings of obscurity, isolation, and fear that originally drove many to misinformation. In line with recent scholarship, we find that searching for truth is often not the primary purpose of (mis)information consumption (Duque and Peres-Neto, 2022; Zimdars et al., 2023). Emotional needs—like desires for belonging, recognition, and control—create strong motivations for consuming and sharing misinformation.

Research shows that fear and lack of control increase susceptibility to misinformation, and our participants described feeling both during elections and health crises (Weeks, 2015). This aligns with findings that emotional needs compound in contexts of uncertainty (e.g., pandemics) and events that prompt communal identification (e.g., elections) (Albertson and Gadarian, 2015).

During the pandemic, sharing (mis)information in reaction to fear provided a sense of purpose, control, and community in extended social isolation (Freiling et al., 2022, 2023).

We found additional emotions affecting misinformation beliefs: obscurity, loneliness, and frustration. We found these emotional states heightened not only by the pandemic but also its aftermath. They were also fueled by political volatility stemming from elections and political disorder. Across both medical and election domains, we find that the impact of such events on participant emotions—and the desire for human connection and community they inculcate—drives consumption, sharing, and creation of misinformation.

We analyzed the consumption and creation of misinformation that meets these needs. Based on our findings, attention to these emotional needs and how misinformation meets them is more important than analyzing how 'truly' individuals believe information and constructing counterarguments. We suggest that alternatively meeting these emotional needs could be more effective in reducing harm than classifying information as false and debunking it.

**Intentionality and Belief**

Researchers have argued that misinformation production implies an element of intent (Guess and Lyons, 2020; Baptista, 2022). Our findings, however, suggest that creators' underlying motivations are rarely singular nor explicitly known by creators themselves. Analyses of the intentionality behind misinformation production must be contextualized in the (often conflicting) emotional, financial, and social needs expressed through online content creation, sharing, and consumption.

While all asserted that they had never intentionally shared misinformation, many of our consumers-turned-creators openly stated that they were partially driven by financial motivation—and articulated that this business acumen helped 'spread the word.' The language and practice of virtuous religious proselytizing and business amplification were overlapping and mutually reinforcing, a phenomenon we believe merits further study.

Further, emotional resonance and identity congruence (Molina, 2023) were powerful trust heuristics for consumers, highlighting the challenge of affecting misinformation beliefs through directly debunking facts or 'neutral' traditional institutional sources of authority.

**Moderation Challenges**

Lastly, this work has implications for platforms' efforts to combat harmful misinformation. Platforms rely on content moderation to reduce misinformation they deem harmful. We observed that many creators anticipate removals or algorithmic efforts to reduce visibility of their posts, preempting these moderation actions by sharing 'gray area' content and migrating to less moderated spaces like Telegram or CloutHub. However, participants rarely made a clean break from old to new platforms; they preferred to maintain a presence on as many platforms as possible, including a mix of moderated and less moderated platforms.

This cross-platform presence and proliferation of 'gray area' content means that moderation alone is insufficient to address misinformation. A complementary approach to removing or reducing misinformation is building resilience to it by teaching people common building blocks and manipulation techniques. Growing research into behavioral and cognitive interventions, from boosts to techno-cognition (e.g. adding friction to technical processes), shows promise in proactively reducing misinformation spread (Kozyreva et al., 2020). Further research is needed to analyze how these interventions apply to misinformation consumers and creators.

## Conclusion

We used ethnographic methods to investigate h*ow and why people encounter, trust, and amplify election and health misinformation.* Participants were more likely to engage with misinformation from creators and channels with less than 100K followers, who used grassroots sources of authority to establish legitimacy. Content creators employed subtle aesthetic and rhetorical techniques to blend their content into the everyday media consumed by participants in their safe spaces, targeting participants' emotional and social needs. Creators used repetition to build trust through repeated exposure to ideas, rather than through single 'red pill' events.

Given the dominance of 'gray area' content in participants' online ecosystems, we recommend further research to identify and counter its effects. Its impact on beliefs and behaviors is poorly understood at scale. We suggest integrating anthropological approaches like ours with psychometric evaluations to better understand how misinformation affects emotions and belief formation in context. Finally, the introduction of generative artificial intelligence is observably affecting misinformation consumption and production. Interdisciplinary research is needed to understand how these emerging technologies affect creators' tactics and consumers' practices when encountering misinformation online.